\documentclass[12pt]{kluwer}

\usepackage{graphicx}

\begin{document}

\begin{opening}
\title{The disc -- jet connection in GRS 1915+105}
\author{Klein-Wolt, M.$^{1}$, Fender, R.P.$^{1}$, Pooley, G. G.$^{2}$, Belloni, T.$^{3}$, Migliari, S.$^{3,4}$, Morgan, E. H.$^{5}$, van der Klis, M.$^{1}$}
\institute{$^{1}$API Amsterdam, $^{2}$MRAO Cambridge, $^{3}$OAB Merate, $^{4}$Univ. Milano, $^{5}$MIT Cambridge} 

\end{opening}

\section{Introduction}

The black hole candidate GRS~1915+105 is known for its extremely complex X-ray behaviour, which is not observed in any other X-ray source. Belloni et al. (2000) succesfully described this behaviour as transitions between three basic spectral states: the soft states A and B, and the hard state C. GRS 1915+105 was discovered in 1994 to be the first galactic source to show superluminal motions of radio-emitting ejecta (Mirabel \& Rodriguez 1994). Besides these big radio flares, the source also shows smaller radio oscillation events with periods around 20-40 minutes (Pooley \& Fender 1997).

\section{Comparing the simultaneous radio and the X-ray observations}

\begin{figure*}[]
\centering{\includegraphics[width=5cm, height=11cm,angle=270]{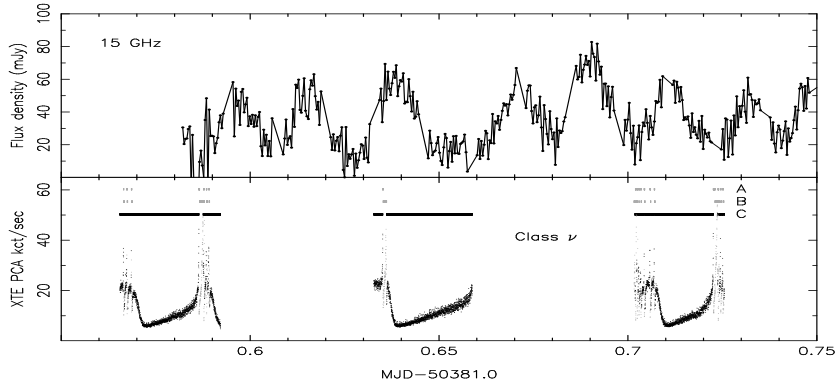}}
\caption{Overlapping Radio and X-ray observation showing the hard X-ray dips (state C), and the radio oscillation events. The rms noise is 3 mJy per point in the radio.}
\label{figs}
\end{figure*}

We have cross-correlated simultaneous (15 GHz) Ryle Telescope (RT) radio and Rossi X-Ray Timing Explorer (RXTE; PCA, 2-60 KeV) X-ray observations of GRS~1915+105 made between June 12, 1996 and December 13, 1999, to find a total of 105 overlapping observations (about 9170 minutes of data, see Klein-Wolt et al. 2000). In Fig.~\ref{figs} we show one representative observation. The X-ray light curve (lower panel) is broken up into states A, B, C. From spectral fits (see Migliari, Vignarca \& Belloni (this volume)) it can be concluded that during the spectrally hard state-C dips the inner part of the accretion disc is either missing or just unobservable. The radio emission, at the same time, shows an oscillating flaring type of behaviour for which the period is found to be between 20 and 40 minutes. We also found observations which do not have state C intervals and consist of state A and B alone. For these observations the inner radius of the accretion disc does not change much and importantly, the radio emission is strongly suppressed: the flux density is typically less than a few mJy, despite up to 50\% variability in the X-ray flux.

We find that radio oscillations require the presence of long ($\geq$100s) state C intervals; for observations consisting of state A and B only the radio emission is suppressed (typically less than a few mJy). With short state C intervals less than 100s in length only very weak radio emission ($\sim$5 mJy) is observed. The long state-C intervals correspond to \emph{hard} dips in the X-ray light curve which seem to have a one-to-one relation with the radio events: each hard dip produces a new radio ``flare''. 

During long ``plateau'' states, when the source is believed to be in state C for up to weeks at a time, a (large) quasi-continuous jet is formed with an almost flat synchrotron spectrum extending to at least the near-infrared. We suggest that both for these large jets and the radio oscillation events shown in Fig.~\ref{figs} the same mechanism could be responsible for the production of the radio emission: during a state C interval the corona and the jet are coupled (or are the same structure) and an outflow takes place. As the material moves downstream in the jet, it becomes optically thin at progressively lower frequencies. This causes the observed delay between the radio and the X-ray emission. The plateau states are just longer and more stable versions of the state C intervals: continuous ejection causes a superposition of optically thick and thin parts to be observed as spectrally flat.

\end{document}